\documentclass[final]{IEEEtran}
\usepackage{amsmath, amssymb}
\usepackage{graphicx}
\usepackage{subfigure}
\usepackage[noadjust]{cite}
\usepackage{color}

\title{Lyapunov Functions Family Approach to Transient Stability Assessment}

\author{Thanh Long Vu and~Konstantin~Turitsyn,~\IEEEmembership{Member,~IEEE}
\thanks{Thanh Long Vu and Konstantin Turitsyn are with the Department of Mechanical Engineering, Massachusetts Institute of Technology, Cambridge, MA, 02139 USA e-mail: longvu@mit.edu and turitsyn@mit.edu.

}}

\begin{document}

 \maketitle
\begin{abstract}
Analysis of transient stability of strongly nonlinear post-fault
dynamics is one of the most computationally challenging parts of
Dynamic Security Assessment. This paper proposes a novel approach
for assessment of transient stability of the system. The approach
generalizes the idea of energy methods, and extends the concept of
energy function to a more general Lyapunov Functions Family (LFF)
constructed via Semi-Definite-Programming techniques. Unlike the
traditional energy function and its variations, the constructed
Lyapunov functions are proven to be decreasing only in a finite
neighborhood of the equilibrium point. However, we show that they
can still certify stability of a broader set of initial conditions
in comparison to the energy function in the closest-UEP method.
Moreover, the certificates of stability can be constructed via a
sequence of convex optimization problems that are tractable even
for large scale systems. We also propose specific algorithms for
adaptation of the Lyapunov functions to specific initial
conditions and demonstrate the effectiveness of the approach on a
number of IEEE test cases.
\end{abstract}


\section{Introduction}

Ensuring secure and stable operation of large scale power systems
exposed to a variety of uncertain stresses, and experiencing
different contingencies are among the most formidable challenges
that power engineers face today. Security and more specifically
stability assessment is an essential element of the decision
making processes that allow secure operation of power grids around
the world. The most straightforward approach to the post-fault
stability assessment problem is based on direct time-domain
simulations of transient dynamics following the contingencies.
Rapid advances in computational hardware made it possible to
perform accurate simulations of large scale systems faster than
real-time \cite{Nagel:2013kf}.

At the same time, the fundamental disadvantage of these approaches
is their overall inefficiency. Reliable operation of the system
implies that most of the contingencies are safe. And certification
of their stability via direct simulations essentially wastes
computational resources. Alternatively, the dynamics following
non-critical scenarios could be proven stable with more advanced
approaches exploiting the knowledge about the mathematical
structure of the dynamic system. In the last decades numerous
techniques for screening and filtering contingencies have been
proposed and deployed in industrial setting. Some of the most
common ideas explored in the field are based on the artificial
intelligence and machine learning approaches
\cite{Wehenkel:1989eo,Fouad:1991bp,He:2013bz,Xu:2012ii}. Most
notable of them is the method of Ensemble Decision Tree Learning
\cite{Diao:2010hu,He:2013bz} that is based on the construction of
hierarchical characterization of the dangerous region in the space
of possible contingencies and operating states.

An alternative set of approaches known under the name of direct
energy methods were proposed in early 80s
\cite{Pai:1981dv,Michel:1983hl,Varaiya1985,Tsolas1985} and
developed to the level of industrial deployments over the last
three decades
\cite{Hiskens1989,Fouad:1991,chang1995direct,Bergen2000,Chiang:2011eo,Hiskens:1997Lya}.
These approaches are based on rigorous analysis of the dynamical
equations and mathematical certification of safety with the help
of the so-called energy functions. Energy functions are a specific
form of Lyapunov functions that guarantee the system convergence
to stable equilibrium points. These methods allow fast screening
of the contingencies while providing mathematically rigorous
certificates of stability. At the same time, limited scalability
and conservativeness of the classical energy methods limits their
applicability and requires enhancement of the method with advanced
algorithms for model reduction. Moreover, the algorithms rely on
identification of unstable equilibrium point (UEP) of energy
function which is known to be an NP-hard problem. In the recent
decades a lot of research was focused on both extension of energy
function to different system components
\cite{ghandhari2001control, chang1995direct} and the improvement
of algorithms that identify the UEPs
\cite{Chiang:1989gn,Liu:1997co,Chen:2009gs}. Remarkably, the
concept of controlling UEP \cite{Chiang:1994ir} provides a
practical and less conservative way to certify stability of the
given fault-cleared state based on knowledge of the fault-on
trajectory.

In this work we extend the ideas of classical energy method and
propose its extension that alleviates some of the drawbacks
discussed above. Basically, this paper makes two main
contributions. First, we show that there exists a convex set of
Lyapunov functions certifying the transient stability of a given
power system, each corresponding to a different stability region
estimate. Second, we introduce an adaptation algorithm to find the
best suited Lyapunov function in the family to specific
contingency situations. The proposed method can generally certify
broader regions of stability compared to the closest-UEP method,
and does not rely on knowledge of the fault-on trajectory as the
controlling-UEP method. Also, the Lyapunov functions family is
constructed via a sequence of Semi Definite Programming (SDP)
problems that are known to be convex. Computational approaches for
solving SDP problems have been in active development in the
mathematical community over the last two decades and were recently
successfully applied to a number of power systems, most
importantly to optimal power flow
\cite{Lavaei:2012fu,Madani:2014hv} and voltage security assessment
\cite{molzahn2013sufficient} problems.

In addition to construction of Lyapunov functions family we
propose several ways of their application to the problem of
certification of power system stability. The first technique
relies on minimization of possibly nonconvex Lyapunov functions
over the flow-out boundary of a polytope in which the Lyapunov
function is decaying. This technique certifies the largest regions
of stability at the expense of reliance on non-convex
optimization. Another alternative is to use only the convex region
of the Lyapunov function, which allows more conservative but fast
certification that can be done with polynomial convex optimization
algorithm. The latter technique is similar to the recently
proposed convex optimizations based on the classical direct energy
method utilized to certify the security of the post-contingency
dynamics \cite{Backhaus:2014}. Finally, as the last alternative we
propose an analytical formulation that does not require any
optimizations at all but also produces conservative stability
certificates.

Applying these stability certificates, we discuss a direct method
for contingency screening through evaluating the introduced
Lyapunov functions at the post-fault state defining the
contingency scenario. Unlike energy function approaches, the
proposed approach provides us with a whole cone of Lyapunov
functions to choose from. This freedom allows the adaptation of
the Lyapunov functions to a specific initial condition or their
family. We propose a simple iterative algorithm that possibly
identifies the Lyapunov function certifying the stability of a
given initial condition after a finite number of iterations.

Among other works that address similar questions we note recent
studies of the synchronization of Kuramoto oscillators that are
applicable to stability analysis of power grids with strongly
overdamped generators \cite{Dorfler:2012, Dorfler:2013}. Also,
conceptually related to our work are recent studies on transient
stability \cite{Anghel:2013} and \cite{caliskan2013compositional}
that propose to utilize network decomposition
 of power grids based on Sum of
Square programming and port-Hamiltonian approach, respectively.

The structure of this paper is as follows. In Section
\ref{sec:stability} we introduce the transient stability problem
addressed in this paper, and reformulate the problem in a
state-space representation that naturally admits construction of
Lyapunov functions. In Section \ref{sec:family} we
explicitly construct the Lyapunov functions and corresponding
transient stability certificates. Section \ref{sec:direct}
explains how these certificates can be used in practice. Finally,
in Section \ref{sec:simulation} we present the results of
simulations for several IEEE example systems. We conclude in
Section \ref{sec:discussion} by discussing the advantages of
different approaches and possible ways in improving the
algorithms.

\section{Transient Stability of Power Systems}\label{sec:stability}

Faults on power lines and other components of power system are the
most common cause for the loss of stability of power system. In a
typical scenario disconnection of a component is followed by the
action of the reclosing system which restores the topology of the
system after a fraction of a second. During this time, however the
system moves away from the pre-fault equilibrium point and
experiences a transient post-fault dynamics after the action of
the recloser.  Similar to other direct method techniques, this
work focuses on the transient post-fault dynamics of the system.
More specifically, the goal of the study is to develop
computationally tractable certificates of transient stability of
the system, i.e. guaranteeing that the system will converge to the
post-fault equilibrium.

In order to address these questions we use a traditional swing
equation dynamic model of a power system, where the loads are
represented by the static impedances and the $n$ generators have
perfect voltage control and are characterized each by the rotor
angle $\delta_k$ and its angular velocity $\dot\delta_k$. When the
losses in the high voltage power grid are ignored the resulting
system of equations can be represented as
\cite{machowski2011power}
\begin{align}\label{eq.swing1}
 m_k \ddot\delta_k + d_k \dot\delta_k + \sum_{j} B_{kj} V_kV_j\sin(\delta_k-\delta_j) - P_k = 0
\end{align}
Here, $m_k$ is the dimensionless moment of inertia of the generator,
$d_k$ is the term representing primary frequency controller action on the governor.
$B_{kj}$ is the $n\times n$ Kron-reduced susceptance matrix with the loads removed from consideration.
$P_k$ is the effective dimensionless mechanical torque acting on the rotor.
The value $V_k$ represents the voltage magnitude at the terminal of the $k^{th}$
generator which is assumed to be constant. 

Note, that more realistic models of power system should include
dynamics of excitation system, losses in the network and dynamic
response of the load. Although we don't consider these effects in
the current work, most of the mathematical techniques exploited in
our work can be naturally extended to more sophisticated models of
power systems. We discuss possible approaches in the end of the
paper.

\begin{figure}[t!]
\centering
\includegraphics[width = 3.2in]{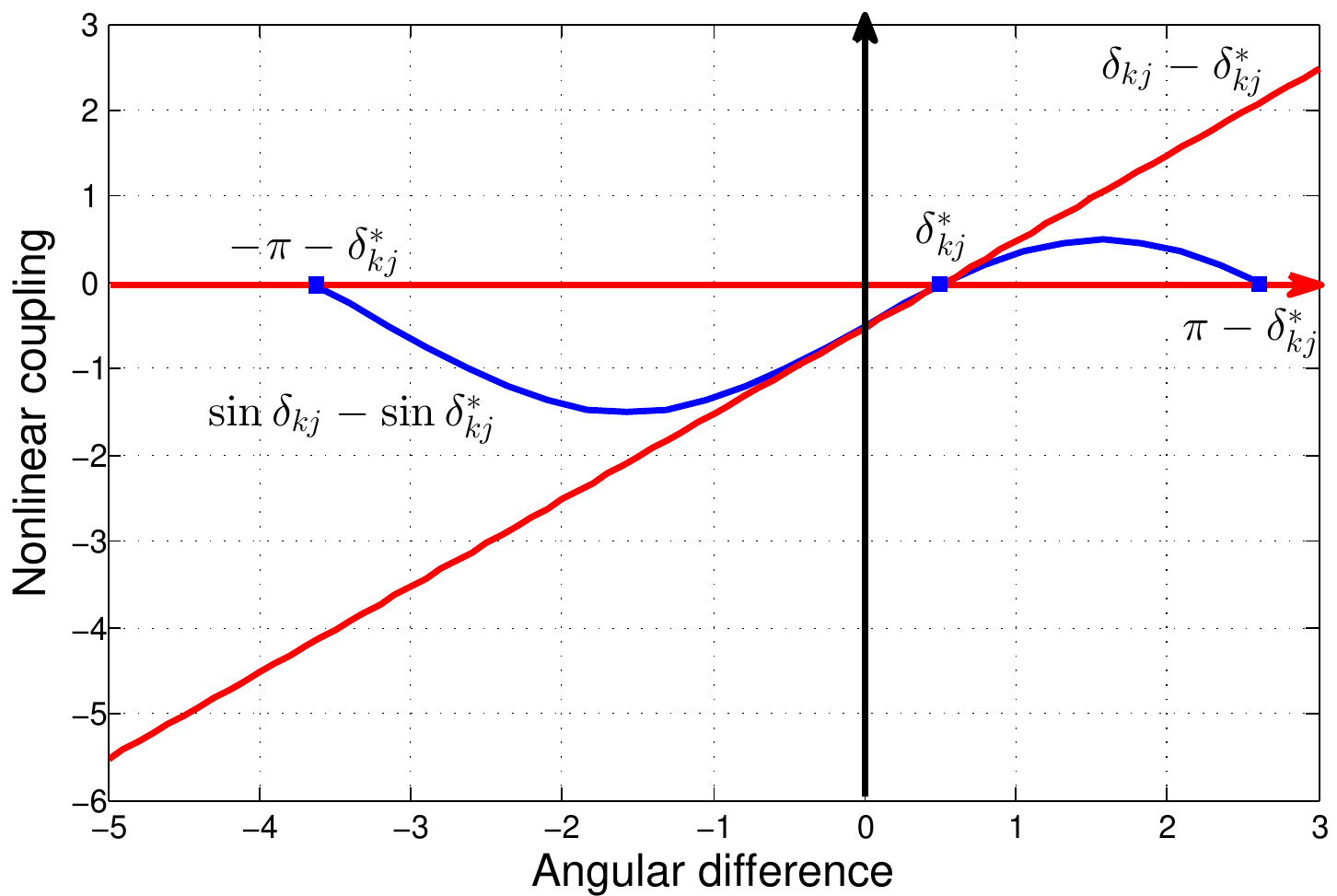}
\caption{Bounding of nonlinear sinusoidal interaction by two
linear functions as described in \eqref{eq.bound}}
\label{fig.NonlinearityBounding}
\end{figure}

In normal operating conditions the system \eqref{eq.swing1} has
many stationary points with at least one stable corresponding to
normal operating point. Mathematically, this point, characterized
by the rotor angles $\delta_k^*$ is not unique, as any uniform
shift of the rotor angles $\delta_k^* \to \delta_k^* + c$ is also
an equilibrium. However, it is unambiguously characterized by the
angle differences $\delta_{kj}^* = \delta_k^*-\delta_j^*$ that
solve the following system of power-flow like equations:
\begin{align}\label{eq.swing2}
 \sum_{j} B_{kj} V_kV_j\sin(\delta_{kj}^*) = P_k
\end{align}
Formally, the goal of our study is to characterize the so called
region of attraction of the equilibrium point $\delta_k^*$, i.e.
the set of initial conditions $\{\delta_k(0),\dot\delta_k(0)\}$
starting from which the system converges to the stable equilibrium
$\delta_k^*$. To accomplish this task we use a sequence of
techniques originating from nonlinear control theory that are most
naturally applied in the state space representation of the system.
Hence, we introduce a state space vector $x = [x_1,x_2]^T$
composed of the vector of angle deviations from equilibrium $x_1 =
[\delta_1 - \delta_1^*\dots \delta_n - \delta_n^*]^T$ and their
angular velocities $x_2 = [\dot\delta_1\dots\dot\delta_n]^T$. In
state space representation the system can be expressed in the
following compact form:
\begin{equation}\label{eq.swing3}
 \dot x = A x - B F(C x),
\end{equation}
with the matrix $A$ given by the following expression:
\begin{align}
A=\left[
        \begin{array}{ccccc}
          O_{n \times n} \qquad & I_{n \times n} \\
          O_{n \times n} \qquad & -M^{-1}D\\
        \end{array}
      \right],
\end{align}
where $M$ and $D$ are the diagonal matrices representing the
inertia and droop control action of the generators, $O_{n\times
n}$ represents the zero and $I_{n \times n}$ the identity matrix
of size $n\times n$. The other matrices in \eqref{eq.swing3} are
given by
\begin{align}
 B= \left[
        \begin{array}{ccccc}
          O_{n \times |\mathcal{E}|} \\
          M^{-1}E^TB \\
        \end{array}
      \right],  C=[E \;\; O_{|\mathcal{E}|\times n}].
\end{align}
Here, $|\mathcal{E}|$ is the number of edges in the graph defined
by the reduced susceptance matrix $B_{kj}$, or equivalently the
number of non-zero non-diagonal entries in $B_{kj}$. $E$ is the
adjacency matrix of the corresponding graph, so that
$E[\delta_1\dots\delta_n]^T =
[(\delta_k-\delta_j)_{\{k,j\}\in\mathcal{E}}]^T$. We assume the
increasing order of $j$ and $k$ for convenience of future
constructions. Finally, the nonlinear transformation $F$ in this
representation is a simple trigonometric function $
F(Cx)=[(\sin\delta_{kj}-\sin\delta^*_{kj})_{\{k,j\}\in\mathcal{E}}]^T.$
The key advantage of this state space representation of the system
is the clear separation of nonlinear terms that are represented as
a ``diagonal'' vector function composed of simple univariate
functions applied to individual vector components. This simplified
representation of nonlinear interactions allows us to naturally
bound the nonlinearity of the system in the spirit of traditional
approaches to nonlinear control
\cite{Popov:1962,Yakubovich:1967,Megretski:1997}. Our Lyapunov
function construction is based on two key observations about the
nonlinear interaction.

First, we observe that for all values of $\delta_{kj} = \delta_k -
\delta_j$  such that $|\delta_{kj} + \delta_{kj}^*| \le \pi$ we
have:
\begin{align} \label{eq.bound}
 0 \le (\delta_{kj}-\delta_{kj}^*)(\sin\delta_{kj} - \sin\delta_{kj}^*) \le (\delta_{kj}-\delta_{kj}^*)^2
\end{align}
This obvious property also illustrated on Fig.
\ref{fig.NonlinearityBounding} allows us to naturally bound the
nonlinear interactions by linear ones. Second, we note that in a
smaller region $|\delta_{kj}|<\pi/2$ the function $\sin\delta_{kj}
- \sin\delta_{kj}^*$ is monotonically increasing, a property that
will play an essential role in proving the convexity of the level
sets of Lyapunov functions in certain regions of the state space.
In the following section we show how these properties of the
system nonlinearity can be used to construct the Lyapunov
functions certifying the transient stability of the system.

\section{Family of Lyapunov functions for stability assessment}\label{sec:family}
The traditional direct method approaches are based on the concept
of the so-called Energy function. The Energy function in its
simplest version is inspired by the mechanical interpretation of
the main equations \eqref{eq.swing1}:
\begin{align} \label{eq.energy}
 E = \sum_k \frac{m_k \dot\delta_k^2}{2} - \sum_{\{k,j\}\in \mathcal{E}}B_{kj}V_kV_j \cos\delta_{kj} - \sum_k P_k \delta_k.
\end{align}
\begin{figure}
\centering
\includegraphics[width = 3.2in]{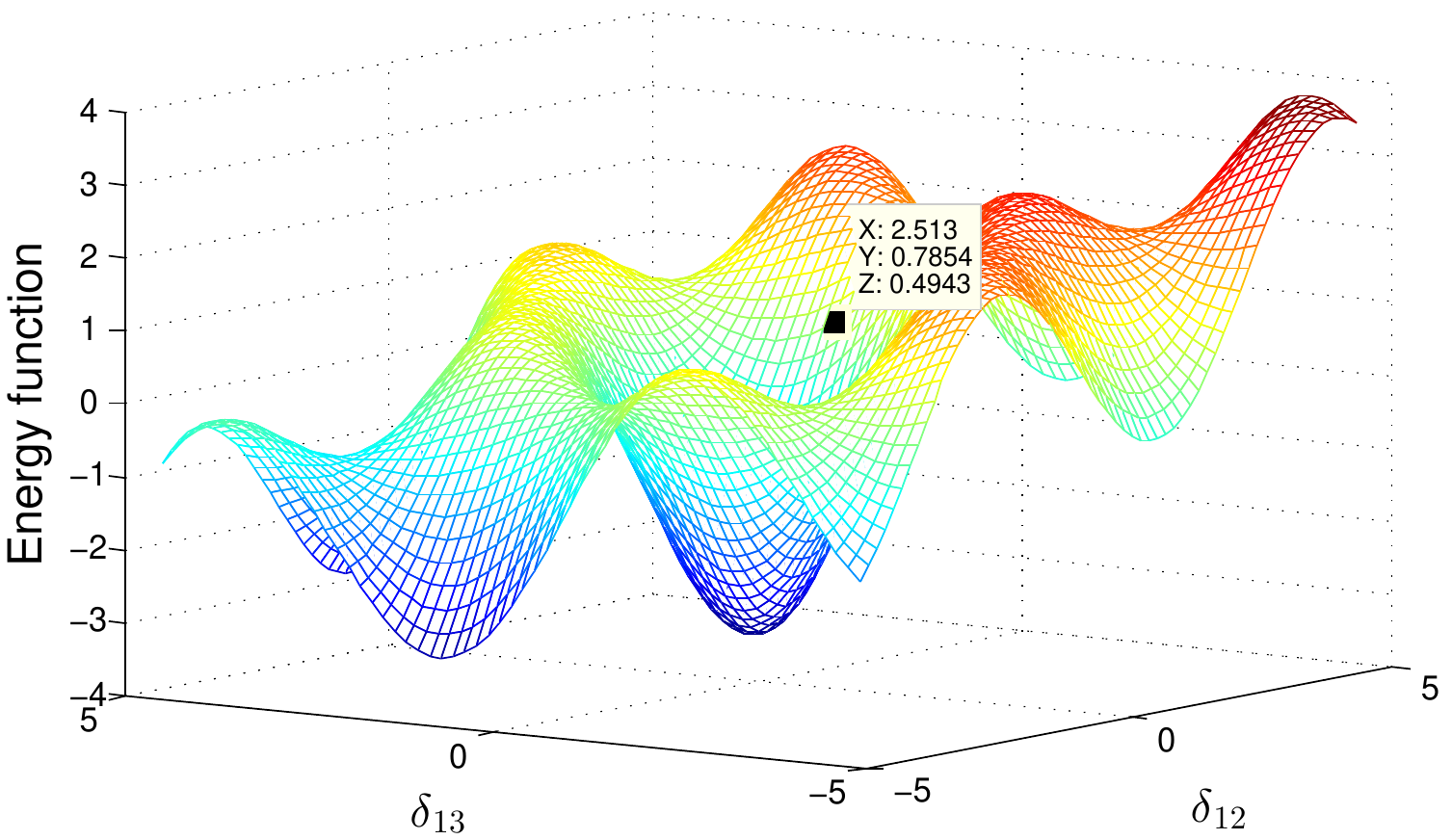}
\caption{Energy function landscape depicted as a projection of the
energy function into the surface defined by the angle differences
$\{\delta_{12}, \delta_{13}\}$} \label{fig.Energy}
\end{figure}
In this expression the first term in the right hand side
represents the kinetic energy of the turbines and the second is
the potential energy of the system stored in the inductive lines
in the power grid network. The dissipative nature of the damping
term in \eqref{eq.swing1} ensures that the energy constructed in
this way is always decreasing in time. Moreover, the energy plays
a role of a Hamiltonian of the system defined for the natural
momentum variables $p_k = m_k \dot\delta_k$, so the conservative
part of the equations of motion \eqref{eq.swing1} can be recovered
via traditional Hamiltonian mechanics approach. This observation
implies, that extrema of the potential energy in \eqref{eq.energy}
are also the equilibrium points of the equations of motion
\eqref{eq.swing1}. An example of Energy function for a simple
$9$-bus system considered in section \ref{sec:simulation} is shown
on Fig. \ref{fig.Energy}. As one can see, the energy function
possesses multiple extrema with only one of them corresponding to
the actual equilibrium point.

Although, the decreasing nature of the energy function provides
the most natural certificate of local stability, it is not the
only function that can be shown to decrease in the vicinity of the
equilibrium point. To illustrate this point qualitatively we first
consider a trivial example of linear dynamics described by the
equation $\dot x = A x$. Whenever matrix $A$ is Hurwitz, the
system has a trivial stable equilibrium $x=0$. Suppose now, that
the left eigenvectors of $A$ are given by $u_k$ respectively, so
that $u_k^T A = \lambda_k u_k^T $, where $\lambda_k$ is the
corresponding eigenvalue. In this case, for every eigenpair there
exists a Lyapunov function defined by $L_k(x) = x^T (u_k
\bar{u_k^T} + \bar{u_k}u_k^T) x \geq 0$, where $\bar{u_k}$
represents the complex conjugate of the vector. This Lyapunov
function is simply the square amplitude of the state projection on
the pair of eigenvectors corresponding to conjugate pair of
eigenvalues. Obviously, as long as the system is stable this
square amplitude is a strictly decaying function. Indeed, one can
check that $d L_k /d t =  2\operatorname{Re}(\lambda_k) L_k \leq 0
$. This construction suggests that any function of type $L(x) =
\sum_k c_k L_k(x)$ with $c_k \geq 0$ is a Lyapunov function
certifying the linear stability of $x^* = 0$. In other words, the
Lyapunov functions of stable linear systems form a simple
orthant-type convex cone defined by inequalities $c_k \geq 0$.

In the context of energy functions, one can interpret the Lyapunov
function $L_k$ as the energy stored in the mode $k$. Obviously for
linear systems, the superposition principle implies that all these
energies are strictly decaying functions. However, in the presence
of nonlinearity, the energy of an individual mode is no longer
strictly decaying, since the nonlinear interactions can transfer
the energy from one mode to another. However, as long as the
effect nonlienarity is relatively small it is possible to bound
the rates of energy transfer and define smaller cone of Lyapunov
functions that certify the stability of an equilibrium point.

For the system defined by \eqref{eq.swing3} we propose to use the
convex cone of Lyapunov functions defined by the following system
of Linear Matrix Inequalities for positive, diagonal matrices $K,
H$ of size $\mathcal{E}\times\mathcal{E}$ and symmetric, positive
matrix $Q$ of size $2n\times 2n$:
\begin{align}
\label{eq.QKH}
    \left[   \begin{array}{ccccc}
          A^TQ+QA  & R \\
          R^T  & -2H\\
        \end{array}\right] &\le 0,
  \end{align}
with $R = QB-C^TH-(KCA)^T$. For every pair $Q,K$ satisfying these
inequalities the corresponding Lyapunov function is given by:
\begin{align} \label{eq.Lyapunov}
V(x) = \frac{1}{2}x^T Q x - \sum_{\{k,j\}\in \mathcal{E}}
K_{\{k,j\}} \left(\cos\delta_{kj}
+\delta_{kj}\sin\delta_{kj}^*\right)
\end{align}
Here, the summation goes over all elements of pair set
$\mathcal{E}$, and $K_{\{k,j\}}$ denotes the diagonal element of
matrix $K$ corresponding to the pair $\{k,j\}$. As one can see,
the algebraic structure of every Lyapunov function is similar to
the energy function \eqref{eq.energy}. The two terms in the
Lyapunov function \eqref{eq.Lyapunov} can be viewed as
generalizations of kinetic and potential energy respectively.
Moreover, the classical Energy function is just one element of the
large cone of all possible Lyapunov functions corresponding to
$K_{\{k,j\}} = B_{kj} V_k V_j$ and $Q$ given by the inertia matrix
$M$.

In Appendix \ref{sec.LyapunovDecrease} we provide the formal proof
of the following central result of the paper. The Lyapunov
function $V(x)$ defined by the equation \eqref{eq.Lyapunov} is
strictly decaying inside the polytope $\mathcal{P}$ defined by the
set of inequalities $|\delta_{kj} + \delta_{kj}^*| < \pi$. This
polytope formally defines the region of the phase space where the
nonlinearity can be bounded from above and below as shown in Eq.
\eqref{eq.bound} and on Fig. \ref{fig.NonlinearityBounding}. In
other words, as long as the trajectory of the system in the state
space stays within the polytope $\mathcal{P}$, the system is
guaranteed to converge to the normal equilibrium point $\delta^*$
where the Lyapunov function acquires its locally minimum value.
The convergence of Lyapunov function is  proved in Appendix
\ref{sec.Convergence} by using the LaSalle's Invariance Principle.
Here, the Lyapunov function $V(x)$ is possibly negative. However,
if we add the constant $\sum_{\{k,j\}\in \mathcal{E}} K_{\{k,j\}}
(\cos\delta_{kj}^*+\delta_{kj}^* \sin\delta_{kj}^*)$ into the
Lyapunov function $V(x),$ we will obtain a function which is
positive definite in $\mathcal{P}$ and whose derivative is
negative semidefinite in $\mathcal{P}.$ Hence, we can rigorously
apply the LaSalle's Invariance Principle.

We note that there may exists the case when the initial state lies
inside $\mathcal{P}$, but after some time periods, the system
trajectory escapes the polytope $\mathcal{P}$, and then the
Lyapunov function is no longer decreasing. In order to ensure that
the system will not escape the polytope $\mathcal{P}$ during
transient dynamics we will add one condition to restrict the set
of initial states inside $\mathcal{P}.$ Formally, we define the
minimization of the function $V(x)$ over the union
$\partial\mathcal{P}^{out}$ of the flow-out boundary segments
$\partial\mathcal{P}_{kj}^{out}$:
\begin{align}\label{eq.Vmin1}
 V_{\min}=\mathop {\min}\limits_{x \in \partial\mathcal{P}^{out}} V(x),
\end{align}
where $\partial\mathcal{P}_{kj}^{out}$ is the flow-out boundary
segment of polytope $\mathcal{P}$ that is defined by $|\delta_{kj}
+\delta_{kj}^*| = \pi$ and $\delta_{kj}\dot{\delta}_{kj} \ge 0$.
Given the value of $V_{\min}$ the invariant set of the Lyapunov
function $V(x)$ where the convergence to equilibrium is certified
is given by
\begin{align}\label{eq.invariant}
 \mathcal{R} = \left\{x \in\mathcal{P}: V(x) < V_{\min}\right\}.
\end{align}
Indeed, the decay property of Lyapunov function in the polytope
$\mathcal{P}$ ensures that the system trajectory cannot meet the
boundary segments $\{x: V(x)=V_{\min}\}$ and
$\partial\mathcal{P}_{kj}^{out}$ of the set $\mathcal{R}.$ Also,
once the system trajectory meets the flow-in boundary segment
$\partial\mathcal{P}_{kj}^{in}$  defined by $|\delta_{kj}
+\delta_{kj}^*| = \pi$ and $\delta_{kj}\dot{\delta}_{kj} < 0,$ it
can only go in the polytope $\mathcal{P}.$ Hence, the set
$\mathcal{R}$ is invariant, and thus, is an estimate of the
stability region.

\begin{figure}
\centering
\includegraphics[width = 3.2in]{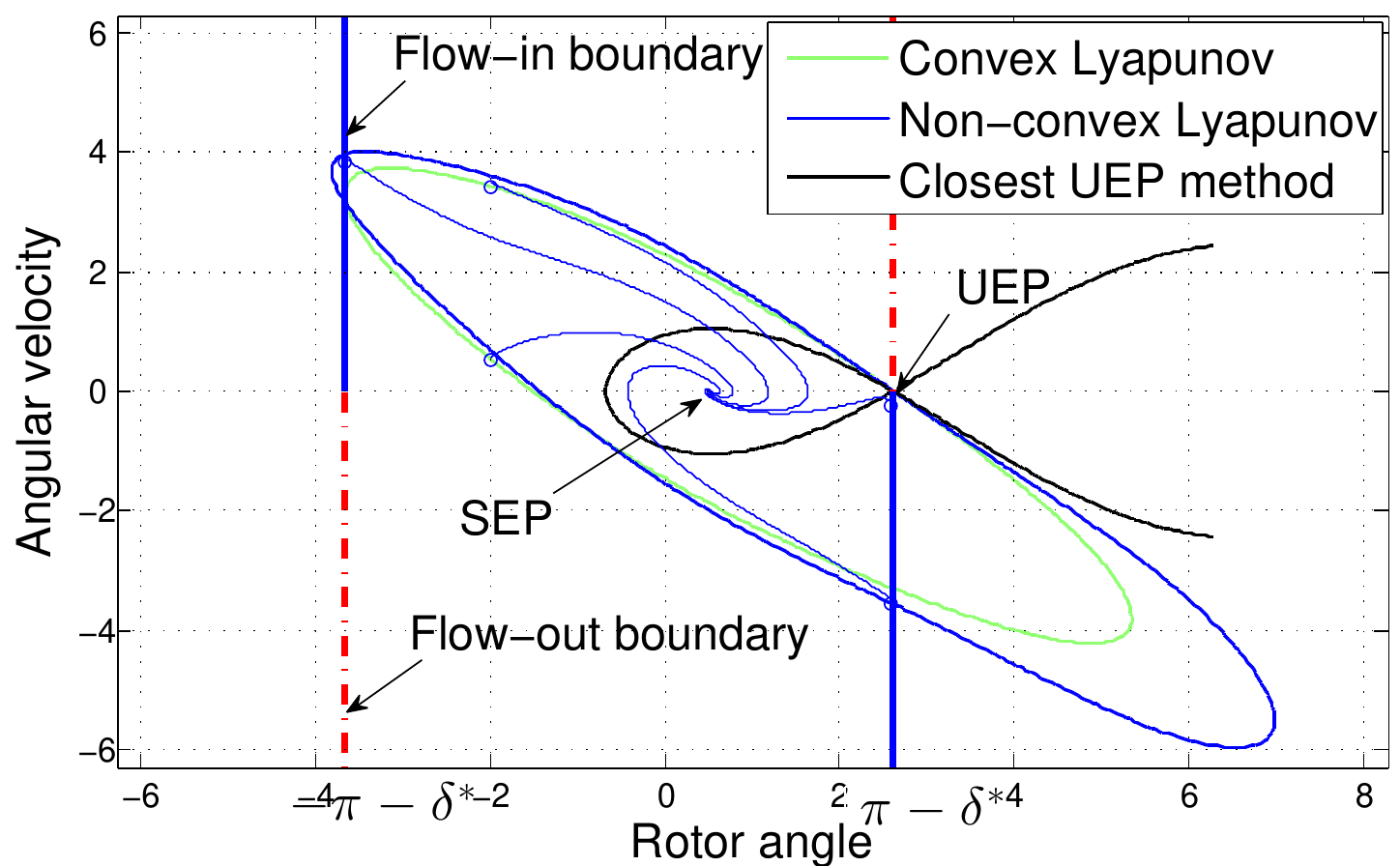}
\caption{Comparison between invariant sets defined by convex and
non-convex Lyapunov functions and the stability region  obtained
by the closest UEP energy method (black solid line). Invariant
sets are intersection of the Lyapunov level sets (blue and green
solid lines) and the polytope defined by $-\pi-\delta^* \le \delta
\le \pi-\delta^*.$} \label{fig.2Bus}
\end{figure}

Note that the stability region estimate is different for different
choice of Lyapunov function. This allows for adaptation of the
certificate to given initial conditions as well as the extension
of the certified set by taking the union of estimates from all the
Lyapunov functions. In the next section we describe the possible
applications of this adaptation technique to the security
assessment problem, while in section \ref{sec.simulations} and on
the Fig. \ref{fig.2Bus} we show that the invariant sets defined by
the Lyapunov functions are generally less conservative in
comparison to the classical Energy method (closest UEP method).

We now discuss techniques to find the value of $V_{\min}$  for a
given choice of Lyapunov function in the family. This task can be
computationally difficult as both the function $V(x)$ and the
boundary of the polytope $\partial \mathcal{P}$ are non-convex. In
order to reduce the complexity of the stability certification we
introduce three constructions of $V_{\min}$ that can be more
computationally tractable, although resulting in more conservative
stability region estimate at the same time. While the first
construction relies on non-convex optimization and results in  the
largest estimate of stability region, the second one only uses
convex region of the Lyapunov function and allows fast but more
conservative certification. Finally, the third construction
proposes an analytical approximation of $V_{\min}$ that does not
require any optimizations but produces conservative stability
certificate.

In particular, the first construction is based on the observation
that $V_{\min}$ can be equivalently defined as the maximum value
at which the largest Lyapunov function's sublevel set does not
intersect the flow-out boundary $\partial\mathcal{P}^{out}$ of the
polytope $\mathcal{P}.$ With each sublevel set $\mathcal{S}(v) =
\{x: V(x) < v\},$ we can find the following maximum value:
\begin{align}
\label{eq.gradient}
g(v) &= \max_{ \{k,j\}\in \mathcal{E},
x \in \mathcal{S}(v) \cap \partial\mathcal{P} } \delta_{kj}\dot{\delta}_{kj}
\end{align}
A sublevel set that does not intersect the flow-out boundary
$\partial\mathcal{P}^{out}$ of the polytope $\mathcal{P}$ is thus
characterized by the inequality $g(v)< 0.$ So, we can formally
define the minimum value $V_{\min}$ as:
\begin{align}
\label{eq.VminEquivalent}
V_{\min}&=\max_{g(v) < 0} v.
\end{align}
Although this formulation may be easier to use in practice in
comparison to the original defined by \eqref{eq.Vmin1}, the
nonlinear constraint makes this problem non-convex, and difficult
to solve for relatively large systems.

The second construction of $V_{\min}$ is based on the observation
that the function $V(x)$ is convex in the polytope $\mathcal{Q}$
defined by the set of inequalities  $|\delta_{kj}| \leq \pi/2$, or
equivalently $\|\delta_{kj}\|_\infty \leq \pi/2$. So, all the
sublevel sets  that do not intersect the flow-out boundary
$\partial\mathcal{Q}^{out}$ of the polytope $\mathcal{Q}$ will
result in an invariant set as long as $\mathcal{Q} \subset
\mathcal{P}$, condition that holds for most of the practically
interesting situations. The convexity of the Lyapunov function
helps us easily compute the maximum value $g^{convex}(v)$ defined
as in
 \eqref{eq.gradient} with $\partial\mathcal{P}$
replaced by $\partial\mathcal{Q}$ (see also \cite{Backhaus:2014}
for the discussion of similar approach applied to the energy
function based methods). Formally, one can then define the
corresponding value of $V_{\min}$ as
\begin{align}\label{eq.convex}
 V_{\min}^{convex} &= \max_{ g^{convex}(v) < 0} v
\end{align}
Therefore, this certificate unlike the formers can be constructed
in polynomial time.

The third construction of $V_{\min}$ is based on a lower
approximation of the minimization  of $V(x)$ taken place over the
flow-out boundary $\partial\mathcal{P}^{out}$ of polytope
$\mathcal{P}.$ In Appendix \ref{sec.LowerApproximation}, we prove
that the minimum value $V^{\{k,j\}}_{\min}$ of $V(x)$ on the
boundary segment $\partial\mathcal{P}_{kj}^{out}$ of the polytope
$\mathcal{P}$ is larger than $
v_{\{k,j\}}=\frac{(\pm\pi-\delta_{kj}^*)^2}{2C_{\{k,j\}}Q^{-1}C^T_{\{k,j\}}}
-K_{\{k,j\}} (\cos(\pm \pi-\delta_{kj}^*)+(\pm \pi -\delta_{kj}^*)
\sin \delta_{kj}^*) -\sum_{\{u,v\} \neq \{k,j\}} K_{\{u,v\}}
(\cos\delta_{uv}^* + \delta_{uv}^* \sin \delta_{uv}^*). $ As such,
the value of $V_{\min}$ can be approximated by
\begin{align}\label{eq.approximate}
V_{\min}^{approx}= \min_{\{k,j\}\in \mathcal{E}} v_{\{k,j\}}
\end{align}
where the minimization takes places over all elements of pair set
$\mathcal{E}.$ This formulation of $V_{\min},$ though
conservative, provides us with a simple certificate to quickly
assess the transient stability of many initial states $x_0,$
especially those near the stable equilibrium point $\delta^*.$


\section{Direct Method for contingency screening}\label{sec:direct}
\label{sec.screen} The LFF approach can be applied to transient
stability assessment problem in the same way as other approaches
based on energy function do. For a given post-fault state
determined by integration or other techniques the value of $V_0 =
V(x_0)$ can be computed by direct application of
\eqref{eq.Lyapunov}. This value should be then compared to the
value of $V_{\min}$ calculated with the help of one of the
approaches outlined in the previous section. Whenever $V_0 <
V_{\min}$ the configuration $x_0$ is certified to converge to the
equilibrium point. If, however $V_0 \ge V_{\min}$, no guarantees
of convergence can be provided but the loss of stability or
convergence to another equilibrium cannot be concluded as well.
These configurations cannot be screened by a given Lyapunov
function and should be assessed with other Lyapunov functions or
other techniques at all.

The optimal choice among three different approaches for
calculation of $V_{\min}$ is largely determined by the available
computational resources. Threshold defined by \eqref{eq.Vmin1}
corresponds to the least conservative invariant set. However, the
main downside of using \eqref{eq.Vmin1} is the lack of efficient
computational techniques that would naturally allow to perform
optimization over the non-convex boundary of the polytope
$\partial \mathcal{P}^{out}$. The second formulation of $V_{\min}$
in \eqref{eq.convex} based on convex optimizations makes it easier
to compute by conventional computation techniques, but results in
a more conservative invariant set. Finally, the third approach
defined by \eqref{eq.approximate} can be evaluated without any
optimizations at all, but also provides more conservative
guarantees.

The main difference of the proposed method with the energy method
based approaches lies in the choice of the Lyapunov function.
Unlike energy based approaches the LFF method provides a whole
cone of Lyapunov functions to choose from. This freedom can be
exploited to choose the Lyapunov function that is best suited for
a given initial condition or their family. In the following we
propose a simple iterative algorithm that identifies the Lyapunov
function that certifies the stability of a given initial condition
$x_0$ whenever such a Lyapunov function exits. The algorithm is
based on the repetition of a sequence of steps described below.

\begin{figure}
\centering
\includegraphics[width = 3.2in]{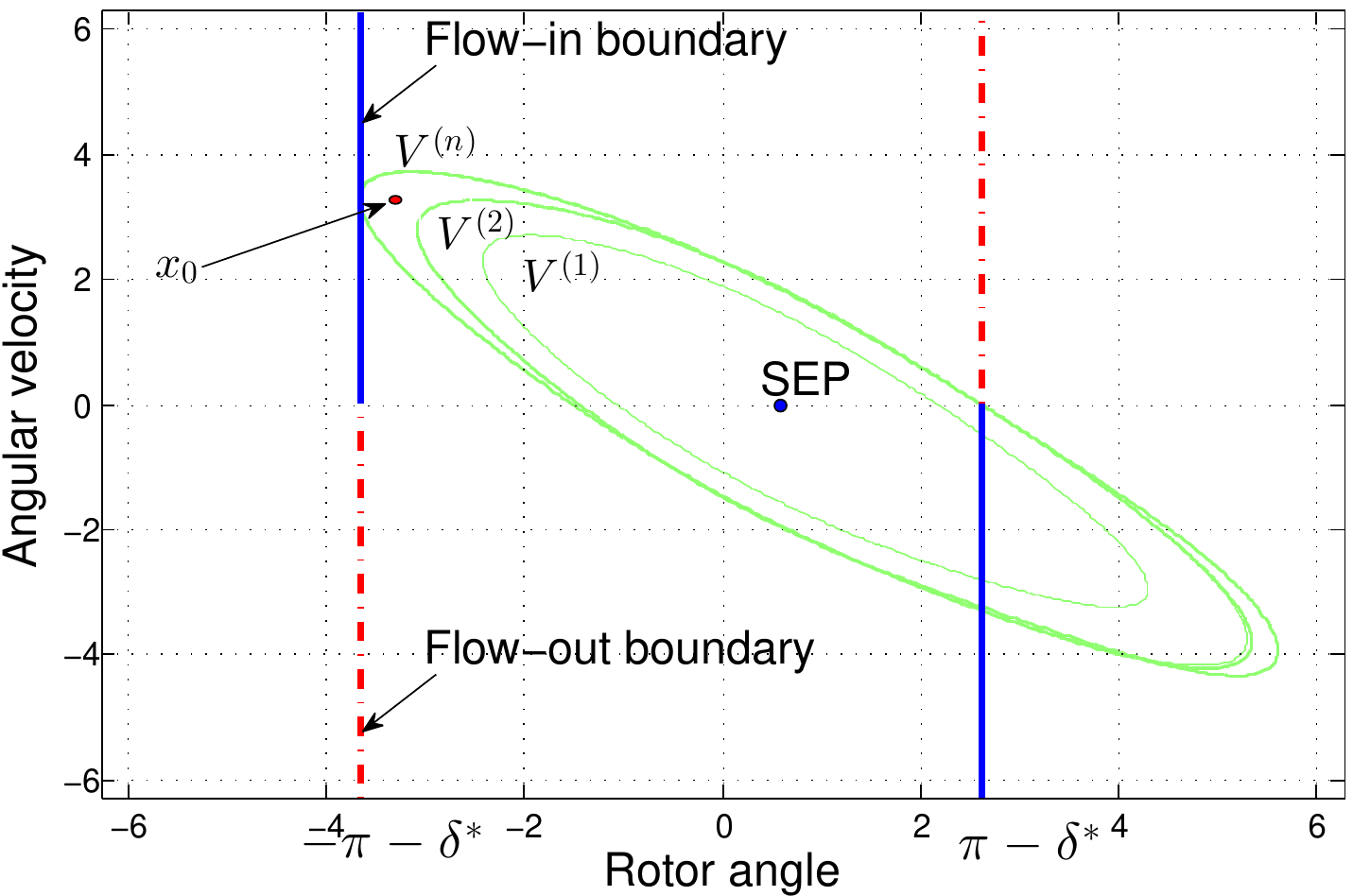}
\caption{Adaptation of the Lyapunov functions to the contingency
scenario over the iterations of the identifying algorithm in
Section \ref{sec.screen}} \label{fig.Adaptation}
\end{figure}
First, we start the algorithm by identifying some Lyapunov
function $V^{(1)}$ satisfying the LMIs \eqref{eq.QKH}, evaluate
the function at the initial condition point $V^{(1)}(x_0),$ and
find the  value of $V_{\min}^{(1)}$. As long as the equilibrium
point is stable such a function is probably guaranteed to exist, one
possible choice would be the traditional energy function. Next, we
solve again the problem \eqref{eq.QKH} with an additional
constraint $V^{(2)}(x_0) < V_{\min}^{(1)} - \epsilon$, where
$\epsilon$ is some step size. Note that the expression
$V^{(2)}(x_0)$ is a linear function of the matrices $Q,K,H$ to
imposing this constraint preserves the linear matrix inequality
structure of the problem. If a solution is found, two alternatives
exit: either $V_{\min}^{(2)} > V^{(2)}(x_0)$ in which case the
certificate is found; or $V_{\min}^{(2)} \leq V^{(2)}(x_0)$ in which case
the iteration is repeated with $V^{(1)}$ replaced by $V^{(2)}$.
In the latter, we have $V^{(2)}_{\min}\leq V^{(2)}(x_0) <  V_{\min}^{(1)} -
\epsilon.$ Hence, the value of $V_{\min}$ is decreasing by at
least $\epsilon$ in each of the iteration step, and thus, the
technique is guaranteed to terminate in a finite number of steps.
Once the problem is infeasible, the value of $\epsilon$ is reduced
by a factor of $2$ until the solution is found. Therefore,
whenever the stability certificate of the  given initial condition
exists it is possibly found in a finite number of iterations.
Figure \ref{fig.Adaptation} illustrates the adaptation of Lyapunov
functions over iterations to the initial states in a simple 2-bus
system considered in Section \ref{sec.simulations}.

\section{Simulation results}\label{sec:simulation}
\label{sec.simulations}
\subsection{Classical 2 bus system}

The effectiveness of the LFF approach can be most naturally
illustrated on a classical $2$-bus with easily visualizable
state-space regions. This system is described by a single 2-nd
order differential equation
\begin{align}
  m \ddot{\delta} +d \dot{\delta} + a \sin\delta - P=0.
\end{align}
For this system $\delta^*=\arcsin(P/a)$ is the only stable
equilibrium point (SEP). For numerical simulations, we choose
$m=1$ p.u., $d=1$ p.u., $a= 0.8$ p.u., $P=0.4$ p.u., and
$\delta^*=\pi/6.$ Figure \ref{fig.2Bus} shows the comparison
between the invariant sets defined by convex and non-convex
Lyapunov functions with the stability region obtained by the
closest UEP energy method. It can be seen that there are many
contingency scenarios defined by the configuration $x_0$ whose
stability property cannot be certified by the closest UEP energy
method, but can be guaranteed by the LFF method. Also, it can be
observed that the non-convex Lyapunov function in \eqref{eq.Vmin1}
provides a less conservative certificate compared to the convex
Lyapunov function, at the price of an additional computational
overhead. For the obtained Lyapunov function, it can be computed
that $V_{\min}=V_{\min}^{approx}=0.7748$ and
$V_{\min}^{convex}=0.2073.$

Figure \ref{fig.Adaptation} shows the adaptation of the Lyapunov
function identified by the algorithm in Section \ref{sec.screen}
to the contingency scenario defined by the initial state $x_0.$ It
can be seen that the algorithm results in Lyapunov functions
providing increasingly large stability regions until we obtain one
stability region containing the initial state $x_0$.

\subsection{Kundur 9 bus 3 generator system}

Next, we consider the 9-bus 3-generator system with data as in
\cite{Anderson:2003}. When the fault is cleared, the post-fault
dynamics of the system is characterized by the data presented in
Tab. \ref{tab.data}.

\begin{table}[ht!]
\centering
\begin{tabular}{|c|c|c|}
  \hline
  Node & V (p.u.) & P (p.u.) \\
  \hline
  1 & 1.0566 & -0.2464 \\
  2 & 1.0502 & 0.2086 \\
  3 & 1.0170 & 0.0378 \\
  \hline
\end{tabular}
\caption{Voltage and mechanical input} \label{tab.data}
\end{table}
\begin{table}[h!]
\centering
\begin{tabular}{|c|c|c|c|}
  \hline
  Node & 1 & 2 & 3\\
  \hline
  1 & 1.181-j2.229 & 0.138+j0.726& 0.191+j1.079 \\
  2 & 0.138+j0.726 & 0.389-j1.953& 0.199+j1.229 \\
  3 & 0.191+j1.079 & 0.199+j1.229& 0.273-j2.342 \\
  \hline
\end{tabular}
\caption{Reduced transmission admittance matrix}
\label{tab.susceptances}
\end{table}

\begin{figure}
\centering \subfigure[Decrease of the Lyapunov function obtained
by the identifying algorithm in Section \ref{sec.screen}]{
\includegraphics[width = 3.2in]{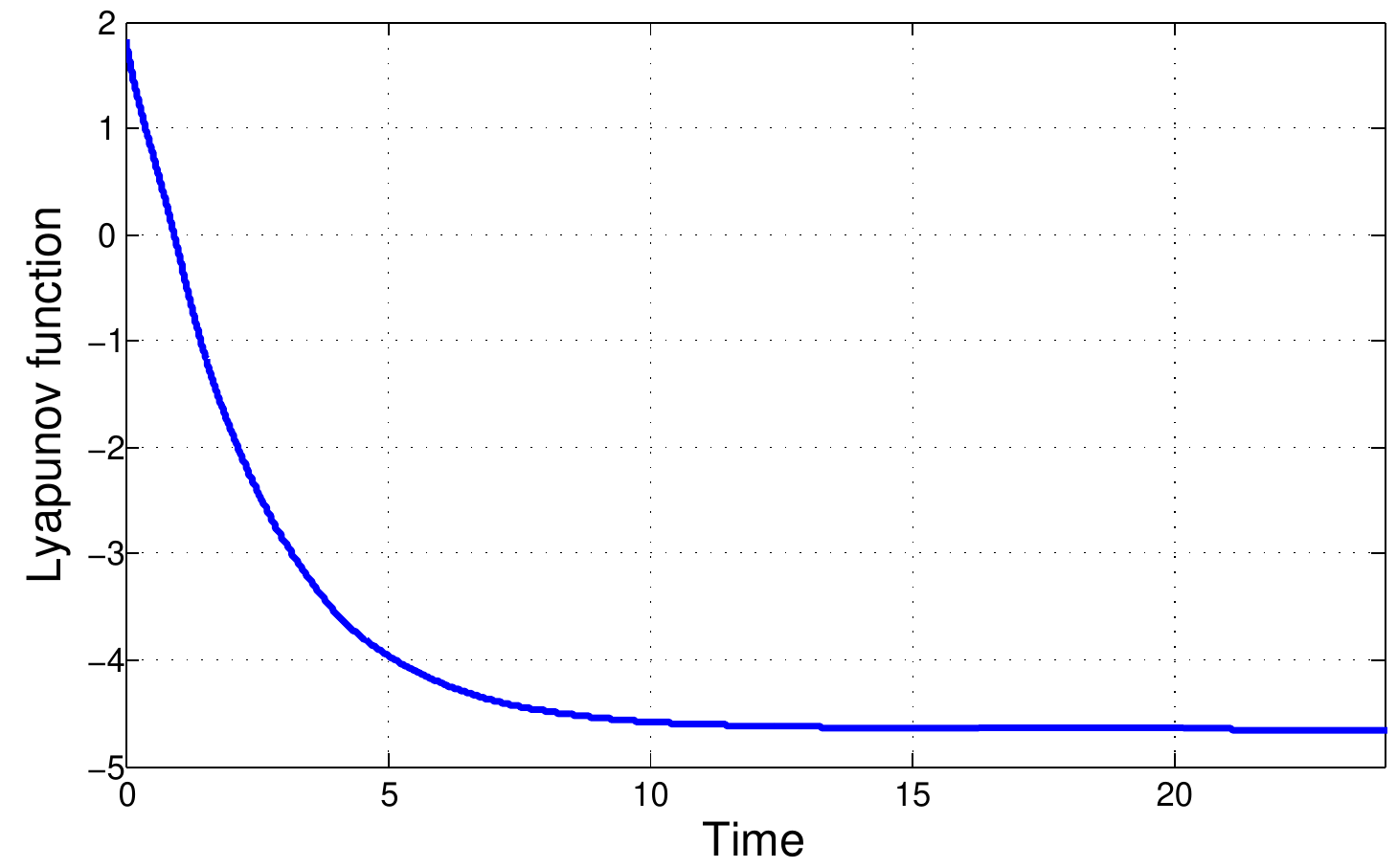}}
 \subfigure[Convergence of generators' angles from the initial state $\{\delta_{12}(0)=2.513, \delta_{13}(0)=0.7854\}$ to the equilibrium $\{\delta_{12}^*=-0.1588, \delta_{13}^*=-0.1005\}$]{
\includegraphics[width = 3.2in]{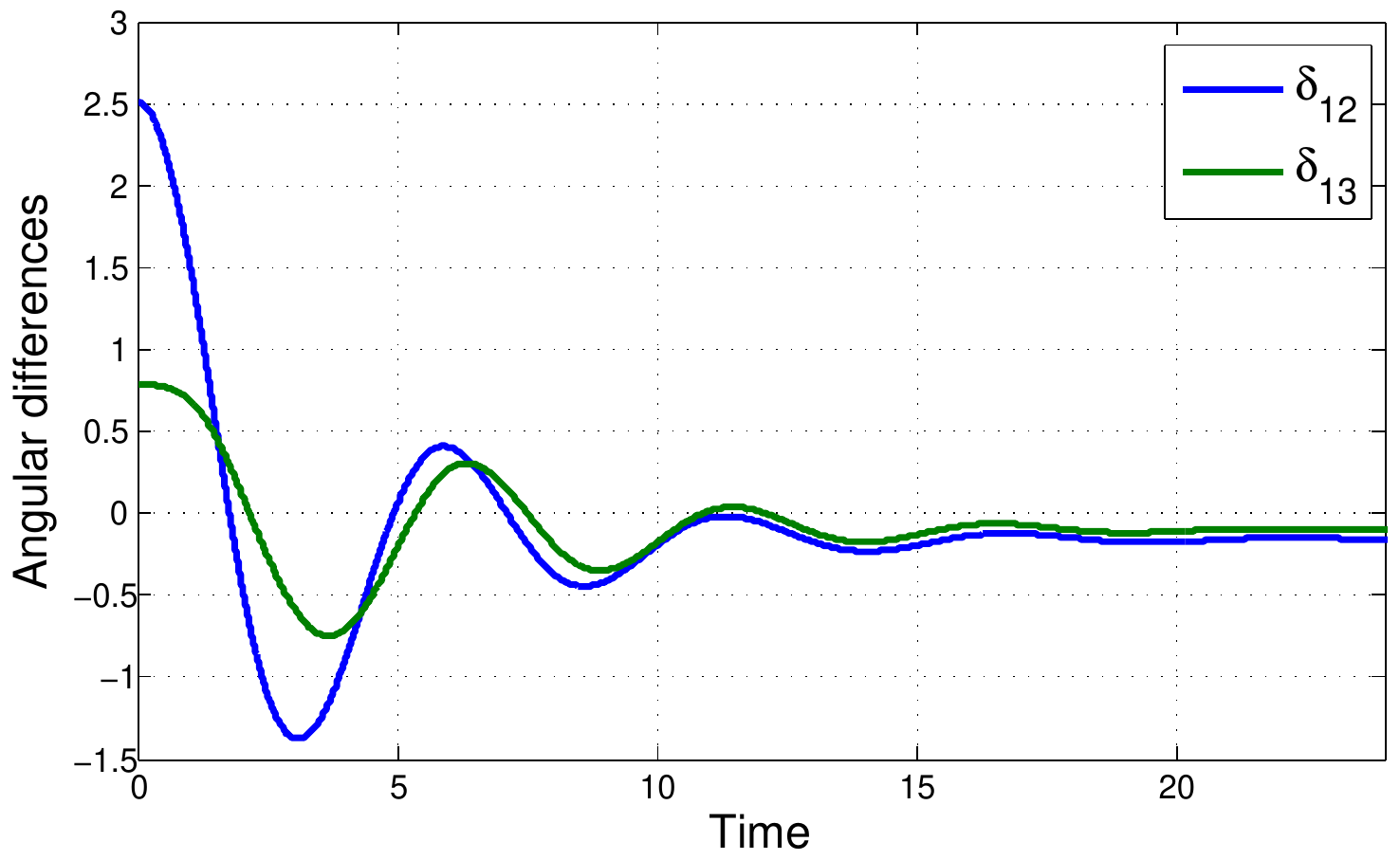}}
 \caption{Post-fault dynamics of a 9 bus 3 generator system}
 \label{fig.9Bus}
\end{figure}

The reduced transmission admittance matrix is given in Tab.
\ref{tab.susceptances}, from which we have, $B_{12}\approx
|Y_{12}|=0.739$ p.u., $B_{13}\approx |Y_{13}|=1.0958$ p.u.,
$B_{23}\approx |Y_{23}|=1.245$ p.u. By \eqref{eq.swing2}, we can
calculate the stable equilibrium point: $\delta_{12}^*=-0.1588,
\delta_{13}^*=-0.1005.$ For simplicity, we take $m_k=2$ p.u.,
$d_k=1$ p.u. Figure \ref{fig.Energy} shows the landscape of the
energy function \eqref{eq.energy}. From Fig. \ref{fig.Energy}, it
can be observed that the stability of the contingency defined by
the initial state $\{\delta_{12}(0)=2.513,
\delta_{13}(0)=0.7854\}$ cannot be guaranteed by the energy method
since the initial energy, $E(0)=0.4943,$ is larger than the
critical energy, which is about $0.196.$ Yet, we can find a
Lyapunov function based on the proposed method that certifies the
stability of contingency defined by the initial state
$\{\delta_{12}(0)=2.513, \delta_{13}(0)=0.7854\},$ as can be
interpreted from the strict decrease of Lyapunov function in Fig.
\ref{fig.9Bus}(a). The convergence of the system from the initial
state $\{\delta_{12}(0)=2.513, \delta_{13}(0)=0.7854\}$ to the
equilibrium point is confirmed by simulation as in Fig.
\ref{fig.9Bus}(b).

\subsection{New England 39 bus 10 generator system}
To illustrate the scalability of the proposed approach, we
consider the New England 39 bus 10 generator system, and evaluate
the construction of Lyapunov function defined by
\eqref{eq.Lyapunov}. The equilibrium point is obtained by solving
the power-flow like equations \eqref{eq.swing2}. The LMIs
\eqref{eq.QKH} are solved by the regular MATLAB sofware CVX to
find the symmetric, positive matrix $Q$ of size $20\times 20$ and
diagonal matrices $K,H$ of size $45 \times 45.$ It takes about
$2.5s$ for a normal laptop to solve these equations, by which the
Lyapunov function $V(x)$ is achieved.

\section{Discussion of the results}\label{sec:discussion}
The Lyapunov Functions Family approach developed in this work is
essentially a generalization of the classical energy method. It is
based on the observation that there are many Lyapunov functions
that can be proven to decay in the neighborhood of the equilibrium
point. Unlike the classical energy function, the decay of these
Lyapunov functions can be certified only in finite region of the
phase space corresponding to bounded differences between the
generator angles, more specifically for the polytope ${\cal P}$
defined by inequalities $|\delta_{ij} + \delta_{ij}^*| < \pi$.
However, these conditions hold for practical purposes. Exceedingly
large angle differences cause high currents on the lines and lead
to activation of protective relays that are not incorporated in
the swing equation model.

The limited region of state space where the Lyapunov function is
guaranteed to decay leads to additional conditions incorporated in
the stability certificates. In order to guarantee the stability
one needs to ensure that the system always stays inside the
polytope ${\cal P}$. We have proposed several approaches that
ensure that this is indeed the case. The most straightforward
approach is to inscribe the largest level set that does not
intersect the flow-out boundary $\partial\mathcal{P}^{out}$ of the
polytope ${\mathcal P}$. This approach provides the least
conservative criterion, however the problem of inscription is
generally NP-hard, similar to the problem of identification of
closest unstable equilibria that needs to be solved in the
traditional energy method. This approach is not expected to scale
well for large scale systems. To address the problem of
scalability we have proposed two alternative techniques, one based
on convex optimization and another on purely algebraic expression
that provide conservative but computationally efficient lower
bounds on $V_{\min}$. Both of the techniques have polynomial
complexity and should be therefore applicable even to large scale
systems.

Our numerical experiments have shown that the LFF approach
establishes certificates that are generally less conservative in
comparison to the closest UEP energy approach and may be
computationally tractable to large scale systems. Furthermore, the
large family of possible Lyapunov functions allows efficient
adaptation of the Lyapunov function to a given set of initial
conditions. Moreover, the computational efficiency of the
procedure allows its application to medium size system models even
on regular laptop computers.

\section{Path Forward}
Although the techniques developed in this work address specific
problem in transient stability assessment, the general strategy
proposed in this work offers  opportunities for development of
computationally fast security assessment tools. We envision that
security assessment where a database of stability and security of
certificates is constructed offline using similar approaches that
adopt the Lyapunov function for most common sets of contingencies.
Although the construction of the certificate may take some
significant time, its application to given system state can be
done nearly instantaneously. So, a database of such certificates
applicable to most common contingencies would allow the operator
to certify security with respect to most common events, and focus
the available computational resources on direct simulations of few
contingencies that cannot be certified this way. At the same time,
an extension of this approach similar to the one the authors have
reported in \cite{VuACC2015} allows to certify that certain
regions in operating conditions space are secure with respect to
common contingencies. This approach offers a path for stability
constrained optimization, as the operation in these safe regions
may be enforced in optimization and planning tools.

There are several ways how the algorithm should be improved before
it is ready for industrial deployment. First practical issue is
the extension of the approach to more realistic models of
generators, loads, and transmission network. Although this work
demonstrated the approach on the simplest possible model of
transient dynamics, there are no technical barriers that would
prevent generalization of the approach. Unlike energy methods, our
Lyapunov function construction does not require that the equations
of motion are reproduced by variations of energy function.
Instead, the algorithm exploits the structure of nonlienarity,
which is confined to individual components interacting via a
linear network. This property holds for all the more complicated
models.

More specifically, incorporation of network losses can be easily
accomplished by a simple shift of the polytope ${\mathcal P}$. Simple
first order dynamic load models can be easily incorporated by
extending the vector of nonlinear interaction function $F$. The
most technically challenging task in extension of the algorithm is
to establish an analogue of the bound \eqref{eq.bound} for
higher-order models of generators and loads. This problem is
closely related to the construction of the Lyapunov function that
certifies the stability of individual generator models. The models
of individual generators although being nonlinear have a
relatively small order, that does not scale with the size of the
system.  Hence Sum-Of-Squares polynomial algebraic geometry
approaches similar to ones exploited in \cite{Anghel:2013} provide
an efficient set of computational tools for bounding complicated
but algebraic nonlinearity. We plan to explore this subject in
the forthcoming works.

The next important question is the robustness of the algorithm to the
uncertainty in system parameters, and initial state. As our
algorithm is based on bound of the nonlinearity, it can
naturally be extended to certify the stability of whole subsets of
equilibrium points and initial post-fault states. Although these
certificates will likely be more conservative, they could be
precomputed offline and later applied to broader range of
operating conditions and contingencies.

Finally, we note that  the proposed algorithms in this paper are
not applicable to give assessment for situations when the
post-fault state is unstable. The extension of LFF method to
certify the transient instability of power systems is a possible
direction in our future research.

\section{Acknowledgements}

This work was partially supported by NSF and MIT/Skoltech and
Masdar initiatives. We thank Hung Nguyen for providing the data
for our simulations and M. Chertkov for sharing the unpublished
preprint \cite{Backhaus:2014}. We thank the anonymous reviewers
for their careful reading of our manuscript and their many
valuable comments and constructive suggestions.

\section{Appendix}
\subsection{Proof of the Lyapunov function decay in the polytope $\mathcal{P}$}
\label{sec.LyapunovDecrease} From  \eqref{eq.QKH}, there
exist matrices $X_{|\mathcal{E}| \times 2n}, Y_{|\mathcal{E}|
\times|\mathcal{E}|}$
  such that $A^TQ+QA =-X^TX, QB-C^TH-(KCA)^T =-X^TY,$ and $-2H = -Y^TY.$
  The derivative of $V(x)$ along \eqref{eq.swing3} is:
  \begin{align}
    &\dot{V}(x) = \frac{1}{2}\dot{x}^TQx+\frac{1}{2}x^TQ\dot{x}-\sum K_{\{k,j\}}(-\sin\delta_{kj}+\sin\delta_{kj}^*)\dot{\delta}_{kj}
    \nonumber \\ &=0.5x^T(A^TQ+QA)x-x^TQBF + F^TKC\dot{x} \nonumber \\
    &=-0.5x^TX^TXx-x^T\big(C^TH+(KCA)^T-X^TY\big)F\nonumber \\&+F^TKC(Ax-BF)
\end{align}
Noting that $CB=0$ and $Y^TY=2H$ yields
\begin{align}
\label{eq.dotV} &\dot{V}(x)=-0.5(Xx-YF)^T(Xx-YF)
 - (Cx-F)^THF \nonumber \\
 & =-0.5(Xx-YF)^T(Xx-YF)
    - \sum H_{\{k,j\}}g_{\{k,j\}},
  \end{align}
  where $g_{\{k,j\}}=\big(\delta_{kj}-\delta_{kj}^*-(\sin\delta_{kj}-\sin\delta_{kj}^*)\big)(\sin\delta_{kj}-\sin\delta_{kj}^*).$
  From Fig. \ref{fig.NonlinearityBounding}, we have $g_{\{k,j\}} \ge 0$
  for any $|\delta_{kj}+\delta_{kj}^*|\le \pi.$ Hence, $\dot{V}(x) \le 0, \forall x \in \mathcal{P},$
  and thus the Lyapunov function $V(x)$ is decaying in  $\mathcal{P}.$

 \subsection{Proof of the system convergence to the stable equilibrium}
  \label{sec.Convergence}
  Consider an initial state $x_0$ in the invariant set $\mathcal{R} \subset \mathcal{P}.$ Then, $\dot{V}(x(t)) \le 0$ for all t.
  By LaSalle's Invariance Principle, we conclude that $x(t)$ converges to the set $\{x:\dot{V}(x)=0\}.$ From \eqref{eq.dotV}, if $\dot{V}(x)=0,$
  then $\delta_{kj}=\delta_{kj}^*$ or $\delta_{kj}=\pm \pi-\delta_{kj}^*$ for all pairs $\{k,j\}.$ Hence, in  the set $\{x:\dot{V}(x)=0\},$
  the nonlinearity $F=0$ and the system \eqref{eq.swing3} becomes $\dot{x}=Ax,$ from which it can be proved that
  $x(t)$ converges to some stationary points.
  Therefore, from  $x_0$ the system converges to the stable equilibrium $\delta^*$
  or to some stationary point $x^*$
  lying on the boundary of  $\mathcal{P}.$  Assume that $x(t)$ converges to some stationary point
  $x^*\in \partial\mathcal{P},$ then $x^*\in \partial\mathcal{P}^{out}$ ($\partial\mathcal{P}^{in}$ does
  not contain any stationary point since $\delta_{kj}\dot{\delta}_{kj} <0, \forall x \in \partial\mathcal{P}^{in}_{kj}$). By definition of $V_{\min}$ and
  $\mathcal{R}$, we have $V(x_0) <V_{\min} \le V(x^*),$
  which is a contradiction with the fact that $V(x(t))$ is decaying in the invariant set $\mathcal{R}.$

  \subsection{Proof of the lower approximation of $V_{min}$}
  \label{sec.LowerApproximation}
\noindent  Let $I_{\{u,v\}}=\cos\delta_{uv}^* + \delta_{uv}^* \sin
\delta_{uv}^*-\cos\delta_{uv} - \delta_{uv} \sin \delta_{uv}^*,$
then
\begin{align}
\label{eq.Vnew}
&V^{\{k,j\}}_{\min} + \sum_{\{u,v\} \neq \{k,j\}} K_{\{u,v\}} (\cos\delta_{uv}^* + \delta_{uv}^* \sin \delta_{uv}^*) \nonumber \\
&= \mathop {\min}\limits_{x \in \partial \mathcal{P}_{kj}^{out}}
\Big[0.5x^TQx -  K_{\{k,j\}} \left(\cos\delta_{kj}
+\delta_{kj}\sin\delta_{kj}^*\right) \nonumber \\ &\qquad
\qquad\qquad+ \sum_{\{u,v\} \neq \{k,j\}} K_{\{u,v\}} I_{\{u,v\}}
\Big],
\end{align}
Note, that $I_{\{u,v\}} \ge 0, \forall x \in \mathcal{P},$ and the
second term in the right hand side of
 \eqref{eq.Vnew} is a constant on $\partial \mathcal{P}_{kj}^{out}.$ Hence,
\begin{align*}
&V^{\{k,j\}}_{\min} + \sum_{\{u,v\} \neq \{k,j\}} K_{\{u,v\}} (\cos\delta_{uv}^* + \delta_{uv}^* \sin \delta_{uv}^*)  \nonumber \\
&\ge \mathop {\min}\limits_{x \in \partial\mathcal{P}_{kj} }(0.5x^TQx)- K_{\{k,j\}} \left(\cos\theta_{kj}^* +\theta_{kj}^*\sin\delta_{kj}^*\right)  \nonumber \\
&= \frac{(\theta_{kj}^*)^2}{2C_{\{k,j\}}Q^{-1}C^T_{\{k,j\}}} -
K_{\{k,j\}} \left(\cos\theta_{kj}^*
+\theta_{kj}^*\sin\delta_{kj}^*\right),
\end{align*}
with $\theta_{kj}^*=\pm\pi-\delta_{kj}^*,$ and thus we obtain
\eqref{eq.approximate}.

\bibliographystyle{IEEEtran}
\bibliography{lff}

\end{document}